\documentstyle[12pt]{article}
\def\be{\begin{eqnarray}}
\def\ee{\end{eqnarray}}
\def\nn{\nonumber}
\def\ds{\displaystyle}
\def\H{{\cal H}}
\def\R{{\cal R}}
\def\t{\mbox{\bf t}}
\def\n{\mbox{\bf n}}
\def\x{\mbox{\rm x}}
\def\y{\mbox{\rm y}}
\def\z{\mbox{\rm z}}
\def\r{\mbox{\rm r}}
\def\I{\mbox{\bf 1}}

\begin{document}
\begin{titlepage}
\begin{flushright}
{\large{\sl UPRF-94-400}}
\end{flushright}
\vskip1.5cm
\begin{center}
{\LARGE\sf Monopole Gauge Fields and Quantum}\\
\medskip
{\LARGE\sf Potentials Induced by the Geometry}\\
\medskip
{\LARGE\sf in Simple Dynamical Systems}\\
\vskip0.8cm
{\Large P.\ Maraner}\\
\smallskip
{\large\sl Dipartimento di Fisica, Universit\`a di Parma,}\\
\smallskip
{\large\sl and INFN, Gruppo collegato di Parma,}\\
\smallskip
{\large\sl  Viale delle Scienze, 43100 Parma, Italy}
\vskip0.6cm
{\large May 1994}
\end{center}
\vskip1cm
\begin{abstract}
A realistic analysis shows that
constraining a quantomechanical system produces the effective dynamics to be
coupled with {\sl abelian/non-abelian gauge fields} and {\sl quantum
potentials} induced by the {\sl intrinsic} and {\sl extrinsic geometrical
properties} of the constraint's surface.
 This phenomenon is observable in the effective rotational motion of some
simple polyatomic molecules. By considering specific examples it is shown that
the effective Hamiltonians for the nuclear rotation of linear and symmetric
top molecules are equivalent to that of a charged system moving in a background
magnetic-monopole field. For spherical top molecules an explicit analytical
expression of a non-abelian monopole-like field is found. Quantum potentials
are also relevant for the description of rotovibrational interactions.
\end{abstract}
\end{titlepage}
\setcounter{page}{2}
\

\section{Introduction}
\

One of the first questions addressed by researchers who first studied the
properties of surfaces embedded in the three-dimensional euclidean space, at
the beginning of nineteenth century, may be formulated more or less in this
terms:

\begin{quote}
 {\sl Is it possible for an hypothetical inhabitant of a surface embedded
in the three-dimensional euclidean space to completely reconstruct the
geometrical features of his world by performing physics experiment?}
\end{quote}

\noindent
At that time physics consisted essentially in Newtonian mechanics. It was soon
realized that the free motion of a particle  on a surface only depends on what
are today called the {\sl intrinsic geometric properties}.
That is, it
results completely insensible to the way in which the surface is embedded in
$R^3$, that is on its {\sl extrinsic geometric properties}. As a typical
example it is very easy to be convinced that classical dynamics on a cone
or on a cylinder in locally undistinguishable from that on a plane.
Once recognizing that constraining a dynamical system corresponds to
reducing the motion from the $n$-dimensional euclidean space $R^n$ to a
submanifold, it appears evident that
the extension of this beautiful results to arbitrary submanifold
of $R^n$ lies at  the heart of the classical theory of constrained system.
 It constitutes the geometrical meaning of d'Alambert's principle.

 Almost two century later we  readdress the same question,
extended to arbitrary submanifold of $R^n$, in the realm of quantum mechanics.
In pursuing this goal it is necessary to overcome the prejudice generated by
the
classical results, that is,  dynamics only depends on constraint's intrinsic
geometry.
On the contrary quantum mechanics is very sensitive on extrinsic geometry,
which appears in the effective constrained dynamics by means of the
coupling with {\sl gauge fields} and {\sl quantum potentials} \cite{MD}.
 The first contribution to this theory has been given by H.\ Jensen and H.\
Koppe \cite{JK} who studied the problem of a quantum particle
constrained on a surface in $R^3$ in a realistic way.
The case of a wire in $R^3$ has also been treated
by many authors \cite{qml} and discussed in  definitive form by
S.\ Takagi and T.\ Tanzawa  \cite{TT}. After many attempts \cite{qmm}
the problem for an arbitrary  submanifold of $R^n$ has been finally solved
independently by K.\ Fujii and N.\ Ogawa \cite{FO} and P.\ M.\ and C.\ Destri
\cite{MD} laying the theoretical foundation of a quantum theory of honolomic
constraint.

Besides the mathematical interest of this theory it is important to
understand whether this results are physically significant, that is whether
geometry induced gauge fields and quantum potentials appearing in the effective
constrained dynamics lead to some observable  phenomenon.  This is the main aim
of this paper. The answer that I will give is somewhat surprising, showing
that effect connected with abelian and non-abelian monopole gauge fields and
quantum potentials already has been  observed in physics at the beginning of
the thirty's \cite{TJD}, that is  at the time of Dirac's paper \cite{Dir}
on magnetic monopole and twentyfive years  before the work in which
Yang and Mills introduced the concept of non-abelian gauge field \cite{YM}.
A similar situation has also been found in the context of geometric phase
\cite{MSW,MB}. The mechanism which produces the coupling with gauge fields in
constrained  quantum mechanics is however distinct from that explored by M.\
V.\ Berry, B.\ Simon and F.\ Wilczek and A.\ Zee \cite{Ber}.
 In this paper I consider simple dynamical models of some polyatomic molecules
giving a new analysis of the {\sl rigid body approximation} allowing the
separation of vibrational and rotational nuclear coordinates. Electronic
degrees of freedom are supposed to be separated by means of the
Born-Oppenheimer
approximation and attention is focused on nuclear motion only. The
{\sl ``rigid body''} is considered in the light of the quantum treatment of
holonomic constraint recently developed in Ref.\cite{MD}, showing how
rotovibrational interactions for symmetrical molecules known as Coriolis
interactions \cite{TJD,Her}, are taken into account in the effective rotational
dynamics by means of the coupling with gauge fields and quantum potentials
induced by the constraint. This illustrates the general geometrical mechanism
producing Coriolis coupling, giving at one time a powerful method to
compute them and an observable example of gauge structures and quantum
potentials induced in constrained quantum mechanics.

The paper is organized as follows.
 In section 2, I address the problem of holonomic constraint in quantum
mechanics by discussing the physical meaning of considering the motion
of a quantum system on a surface in $R^3$ or more generally on a submanifold
of $R^n$. I do this by considering a specific example. The fundamental
difference between constrained classical and quantum mechanics is then
outlined and a general definition of a constrained quantum system is given in
terms of a  microscopic potential. The effective quantum dynamics on the
constraint is discussed in section 3. After adapting coordinates and
introducing
the geometrical quantities characterizing the constraint's surface, it is shown
how the general hypothesis on the microscopic nature of the constraint allow to
set up a perturbative description of the motion of the system. The effective
quantum dynamics results covariant in character and depends on extrinsic
geometry by means of the coupling with gauge fields and quantum potentials.

 The theory is applied to simple molecular systems in sections 4,5,6 and 7.
Exploring the {\sl rigid body model} of some polyatomic molecules in the light
of the theory developed in section 3, it is shown how monopole gauge fields
and quantum potentials induced by the {\sl rigidity constraint} are
responsible for long studied rotovibrational interactions.

\

\section{Constraints in Quantum Mechanics}
\

Among the topics which have most attracted the interest of physicists in the
last decade the study of bidimensional systems plays certainly an important
role. Although in many cases the two dimensions are considered in toy models,
in others we are faced with real systems whose dynamics appears to be
effectively
bidimensional. A very important example is found in the description of Quantum
Hall Effect, explained by means  of the properties of a bidimensional electron
gas \cite{Pra}. Let us consider the physical mechanism realizing planarity.
 Quantum Hall Effect is observed in  inversion layers in presence of a strong
magnetic field and at very low temperatures.
Inversion layers, in which the planarity constraint is realized, are formed at
the interface between a semiconductor and an insulator. When a potential
difference is applied an electric field normal to the interface
attracts the free electrons from the semiconductor to the interface itself.
The motion normal to the interface is then confined in a potential
well while the motion in other directions may be considered
free.
In a free electron approximation, relevant for Integer Quantum
Hall Effect, the dynamics separates and the energy of each electron may be
decomposed in a term relative to the  normal motion, $E^{nor}$, and one
relative to the motion  along  the interface, $E^{tan}$,
\be
E=E^{nor}+E^{tan}. \nn
\ee
The  deeper the potential well confining the particle the greater $E^{nor}$
with respect to $E^{tan}$. The spectrum of the system separates in bands.
In correspondence of
every normal eigenstate the whole spectrum of the motion along the interface
may
be observed. The ideal situation to investigate the Quantum Hall Effect is that
in which only the first band is occupied and it is indeed possible to arrange
the experimental apparatus in such a way.  In any physical realization of
the planarity  constraint it is however necessary  to take into account
the whole band spectrum.
 In reducing the motion of a quantum system from $R^3$ to the plane $R^2$ is
therefore
possible to separate the degrees of freedom normal to the constraint to
that along it obtaining an effective bidimensional dynamics. The whole spectrum
keeps nevertheless memory of the way this has been done. The energies are in
first approximation those corresponding to the interface's potential
$V_{I}$ and different shapes of $V_{I}$ produces different band structures.

 The analysis of this simple example shades light on the fundamental difference
between the classical and quantum nature of constraints. Whereas the
classical system may be completely squeezed on the constraint's surface
ignoring motion in normal directions, Heisemberg's principle forbids this
operation in  the quantum case.
The more the system is squeezed on the constraint's surface the more relevant
the motion  in normal directions.
 In describing a quantum system we can't therefore ignore the {\sl physical
mechanism realizing a constraint}, that is the explicit form of the confining
potential $V_{C}$.
The example of inversion layers suggest that two general properties have to
be satisfied by $V_{C}$

\begin{description}
\item{$C1$)} $V_{C}$ {\sl presents a deep minimum in correspondence of the
constraint's surface}
\item{$C2$)} $V_{C}$ {\sl depends only on coordinates normal to the
constraint's surface}
\end{description}

 In next section I will show how  is possible to separate the motion along
and normal to a generic constraint and how the induced dynamics depends on the
explicit form of $V_{C}$.

\

\section{Effective Quantum Dynamics}
\

As usual,
 in addressing the solution of a physical problem the choice of coordinates
adapted to the system under consideration is of
fundamental importance. In the sequel we consider an arbitrary $m$-dimensional
submanifold $M$ of the euclidean space $R^n$. Everywhere the attribute tangent
and normal  are referred to $M$. To adapt coordinates we introduce
\begin{description}
\item{---} a coordinate frame $\{x^\mu; \mu =1,...,m\}$ on $M$, or,
equivalently a smooth assignation of $m$ independent tangent vectors
$\t_\mu(x)$, $\mu =1,...,m$,
\item{---} a choice of $(n-m)$ smooth varying orthonormal normal vectors
$\n^i(x)$, $i=m+1,...,n$, and associated to it, the distances
$\{y^i; i=m+1,...,n\}$ along the geodetics leaving $M$ with speed $\n^i$.
\end{description}
The set $\{x^\mu, y^i; \mu =1,...,m,i=m+1,...,n\}$ constitutes a good
coordinate system for $R^n$ at least in a sufficiently small neighbourhood of
$M$.
 The submanifold $M$ may be completely characterized in terms of the quantities
\cite{Spi}
\be
&g_{\mu\nu}=\t_\mu\cdot\t_\nu &\mbox{{\sl metric (first fundamental
form)}}\nn\\
&\alpha^i_{\mu\nu} = \t_\mu\cdot\partial_\nu\n^i &
                                       \mbox{{\sl second fundamental
form}}\nn\\
&A_\mu^{ij} =\n^i\cdot\partial_\mu \n^j&\mbox{{\sl normal fundamental form}}\nn
\ee
where $\partial_\mu={\partial\over\partial x^\mu}$ and the dot denotes the
scalar product in $R^n$. $g_{\mu\nu}$ describes the intrinsic properties of $M$
whereas $\alpha^i_{\mu\nu}$ and $A_\mu^{ij}$ its extrinsic  geometry. As an
example consider
 an arc-length parameterized curve embedded in $R^3$ with
$\n^2$, $\n^3$  chosen as normal and binormal. Then we obtain : $g_{11}=1$,
reflecting the fact that intrinsic geometry of a one-dimensional manifold
is always trivial; $\alpha^2_{11}=k$, the curve's curvature,  $\alpha^3_{11}=0$
and $A_1^{23}=-A_1^{32}=\tau$, the curve's torsion.

 The metric $G_{IJ}$ of $R^n$ in the {\sl adapted coordinates} frame $\{x^\mu,
y^i\}$ is written solely in terms of $g_{\mu\nu}$,  $\alpha^i_{\mu\nu}$ and
$A_\mu^{ij}$ as \cite{MD}
\be
G_{IJ} = \pmatrix{
    \gamma_{\mu\nu}+y^k y^l A_{\mu}^{kh} A_{\nu}^{lh}&
       y^k A_{\mu}^{jk} \cr
       y^k A_{\nu}^{ik}  & \delta^{ij}},
\ee
where $ \gamma_{\mu\nu} =
g_{\mu\nu}-2y^k\alpha^k_{\mu\nu}+y^ky^l\alpha^k_{\mu\rho}g^{\rho\sigma}
\alpha^l_{\sigma\nu}$. It is important to note that the variation of the
$\n^i(x)$'s by a point dependent rotation $\R^{kl}(x)$ produces $A_\mu^{ij}$ to
transform as an $SO(n-m)$ gauge potential
\be
A_\mu^{ij}\longrightarrow \R^{ik}A_\mu^{kl}\R^{jl}+\R^{ik}\partial_\mu\R^{jk}.
\nn
\ee

 The characterization of a constraint by means of a potential satisfying  the
general conditions given in the previous section and the choice of adapted
coordinates allows to give a complete description of the constrained dynamics.
To achieve this aim
\begin{description}
\item{---}We start by unambiguously quantizing the system in a cartesian
coordinate frame $\{r^I; I=1,...,n\}$ of $R^n$. The Hamiltonian
reads\footnote{In this section $\hbar=1$ and masses are supposed to be
reabsorbed in the relative coordinates.}
\be
\H =-{1\over 2}\sum_{I=1}^{n}{\partial^2\over{\partial r^I}^2} +V_{C}.
\ee

\item{---} Then we transform to adapted coordinates obtaining the Hamiltonian's
expression
\be
\H=-{1\over 2G^{1/2}}\partial_IG^{IJ}G^{1/2}\partial_J +V_{C},
\ee
where $G$ and $G^{IJ}$ denote respectively the determinant an the inverse of
the metric tensor $G_{IJ}$. For a review of quantum mechanics in an arbitrary
coordinates frame see Ref.\cite{Dow}. Everywhere in this paper the sum over
repeated index is understood.

\item{---} Performing the similitude transformation
\be
\H\longrightarrow {G^{1/4}\over g^{1/4}}\H {g^{1/4}\over G^{1/4}}
\ee
we obtain an Hamiltonian acting on wavefunction which are correctly
normalized on the submanifold $M$.

\item{---}  At this point, and only at this point, we implement conditions $C1$
and $C2$ by expanding $V_{C}$ around its minimum
\be
V_{C} = {1\over 2\epsilon^2}{\omega^i}^2{y^i}^2 + a_{ijk}y^iy^jy^k
+b_{ijkl}y^iy^jy^ky^l + ... \ ,
\ee
where the scale of the frequencies $\omega^i$ has been reabsorbed in the
adimensional parameter $\epsilon^{-1}$. The smaller $\epsilon$ the more the
system is squeezed on the constraint's surface.

\item{---} $\epsilon$ appears as a natural perturbative parameter in the
theory.
Rescaling the normal coordinates by ${\vec y} \rightarrow \epsilon^{1/2}
{\vec y}$ the Hamiltonian may be expanded in powers of $\epsilon$ as
\be
&\epsilon\H =& H^{(0)}+\epsilon H^{(1)} +\epsilon^{3/2} H^{(3/2)}+ ... + \nn \\
& & +\epsilon^{5/2} a_{ijk}y^iy^jy^k+\epsilon^3 b_{ijkl}y^iy^jy^ky^l + ...\ \ .
\label{exp}
\ee
\end{description}

 As for the planarity constraint discussed in the second section the zero
order Hamiltonian $H^{(0)}$ depends only on normal degrees of freedom. It
describes a system of $(n-m)$ uncoupled harmonic oscillators with frequencies
$\omega^{m+1},...,\omega^{n}$,
\be
H^{(0)}={1\over 2}\left(-\partial_i\partial_i+{\omega^i}^2 {y^i}^2\right).
\ee
The first order terms $H^{(1)}$, to which I will restrict the analysis in this
paper,
describes the effective dynamics on the constraint. It is covariant in
character
and coupled to the normal dynamics by means of the minimal interaction with a
geometry induced gauge field and a quantum potential
\be
  H^{(1)}=-{1\over 2 g^{1/2}}
   \left( \partial_\mu+{i\over 2}A_\mu^{ij}L_{ij}\right)
   g^{\mu\nu}g^{1/2}
   \left( \partial_\nu+{i\over 2}A_\nu^{kl}L_{kl}\right)
   +Q(x)
\ee
where $L_{ij}=-i(y^i\partial_j-y^j\partial_i)$ are angular momentum operators
in the normal directions and the potential $Q(x)$ may be expressed in terms of
the intrinsic scalar curvature $R$ and the extrinsic mean curvature $\eta$
of $M$ as
\be
Q(x)={1\over 4} R(x) - {m^2\over 8} \eta^2(x).
\ee
Second, third and following terms of the $V_{C}$'s expansion may be of the same
order than $H^{(1)}$ and have eventually to be considered together in
addressing
the perturbation theory.
Higher orders terms, $H^{(3/2)}$, $H^{(2)}$, ... describe the interactions
between  normal and tangent degrees of freedom \cite{Mar}.

 To separate definitively the  dynamics on $M$ from that normal to $M$ we
proceed by means of standard perturbation theory. Let us label with
$E^{(0)}=\sum_i \omega^i(n_i+1/2)$ the, possibly degenerate, states of
$H^{(0)}$. In correspondence of every $E^{(0)}$ the  corrections to the
spectrum have to be evaluated by diagonalizing the perturbation, that is by
solving the Schr\"odinger equation corresponding to Hamiltonian
\be
  \H^{E^{(0)}}=-{1\over 2 g^{1/2}}
   \left( \partial_\mu+ i A_\mu\right)
   g^{\mu\nu}g^{1/2}
   \left( \partial_\nu+ i A_\nu\right)
   +Q(x) +{\bar Q}(x)
\label{Heff}
\ee
with
\be
& & A_\mu = {1\over 2}A_\mu^{ij}\langle L_{ij} \rangle,\label{gf}  \\
& & {\bar Q}(x) ={1\over 8}g^{\mu\nu}A_{\mu}^{ij}A_{\nu}^{kl}
\left(\langle L_{ij}L_{kl}\rangle-
\langle L_{ij}\rangle\langle L_{kl}\rangle\right),\label{Qb}
\ee
where $\langle L_{ij}\rangle$ and $\langle L_{ij}L_{kl}\rangle$ denote the
matrices obtained by braketing $L_{ij}$ and $L_{ij}L_{kl}$ between the
eigenstates corresponding to $E^{(0)}$. $\H^{E^{(0)}}$ describe the effective
motion on the submanifold when the normal degrees of freedom are frozen in the
state $E^{(0)}$. $\langle L_{ij}\rangle$ results
different from zero if and only if $V_{C}$ posses degenerate frequencies, this
being the condition for the coupling with gauge fields not to disappear.
Moreover, if a frequency has degeneracy $d$ the $\langle L_{ij}\rangle$'s with
$i$ and $j$ corresponding to that frequency form a matrix representation of the
generators of $SO(d)$. Denoted by $\omega_1,...,\omega_r$ the distinct proper
frequencies of $V_{C}$ and by $d_1,...,d_r$, $d_1+...+d_r=n-m$, their
degeneracies, the geometry induced gauge group of the theory is
$SO(d_1)\times ...\times SO(d_r)$.

 As a particular case we may recover the Jensen-Koppe answer to the question
proposed in the introduction. The Hamiltonian describing the effective motion
of a quantum particle constrained on a surface embedded in $R^3$ is given by
\be
\H=-{1\over 2}\triangle -{1\over 4}\left({1\over R_1}-{1\over R_2}\right)^2
\ee
where $\triangle$ is the Laplacian on the surface and $R_1$ and $R_2$ its
principal radii of curvature. In this simple example
the dynamics does not  depends on the normal state and the coupling
with geometry induced gauge fields disappears. The drastic difference with
the classical case may nevertheless be observed by considering the motion on
a cone. Whereas the classical particle behaves as free the quantum particle
is attracted on the vertex from a quantum potential \cite{IN}.
 The simpler case in which dynamics results coupled with geometry induced gauge
fields is the wire embedded in $R^3$ as discussed by S.\ Takagi and T.\
Tanzawa \cite{TT}.

 A realistic analysis of constrained quantum systems shows the emergence of a
rich dynamical structure unexpected from semiclassical considerations and
it is a wonderful surprise to rediscover abelian an non-abelian gauge
structures
in a so simple dynamical context. This phenomenon very much resemble Berry's
analysis of the adiabatic approximation \cite{GP} and considering the motion on
the
constraint by freezing normal degrees of freedom is an adiabatic approximation
after all. Nevertheless the mechanism producing the coupling with gauge fields
in constrained quantum mechanics is different from that considered
by M.\ V.\ Berry. In the context of Berry's geometric phase  the embedding of
the slow
coordinates space in the total configuration space is trivial, whereas this
non-triviality lies at the heart of the mechanism appearing in the treatment
of constraint.

\

\section{Molecules as Constrained Systems}
\

Extremely important examples of constrained systems in quantum mechanics are
given by molecular physics. The description of molecular spectra \cite{Her}
starts from the many-body Hamiltonian given by the sum of electronic and
nuclear
kinetic energy plus the coulomb interaction between all the particles. At this
point electronic and nuclear degrees of freedom are separated by means of the
Born-Oppenheimer approximation and an effective Hamiltonian describing the
nuclear motion is obtained
\be
 {\cal H}_{nuc}=-\sum_{a=1}^{N}{\hbar^2\over 2 m_a}
  {\partial^2\over\partial {\vec{r}_a}^2}+V_{BO},
\label{Hnuc}
\ee
where $\vec{r}_a$, $a=1,...,N$,
denotes the coordinates of the $a-$th nucleus, $m_a$ its mass, and $V_{BO}$ the
Born-Oppenheimer potential. For sake of clearness I will ignore throughout
this paper eventual effects connected to the geometric phase. $V_{BO}$
presents a deep minimum in correspondence of the equilibrium configuration
of the molecule. Since this  is determined up to orientation in space,
that is, up to an $SO(3)$ rotation, the previous sentence may be rephrased by
saying
\begin{description}
\item{---} $V_{BO}$ presents a deep minimum in correspondence of the
submanifold $SO(3)$ of the nuclear configuration space $R^{3N-3}$.
\end{description}
Nuclear positions are referred to the body center frame.
Depending only on relative nuclear distances, $V_{BO}$ does not depend on
molecule's orientation in space, that is to say
\begin{description}
\item{---} $V_{BO}$ depends only on coordinates normal to the submanifold
$SO(3)$ of the nuclear configuration space $R^{3N-3}$.
\end{description}
These are exactly conditions $C1$ and $C2$ introduced in the second section to
describe a potential realizing a constraint. The Born-Oppenheimer potential
$V_{BO}$ realizes therefore a {\sl rigidity constraint}\footnote{In the least
decade there has been a growing interest in the study of non-rigid molecules
\cite{Be}.  Following considerations do not apply to these systems.} confining
the nuclear  motion from the configuration space $R^{3N-3}$ to the rotational
group $SO(3)$.

 The use a rigid body model to separate vibrational and rotational degrees of
freedom is a standard matter \cite{Her}. Nevertheless the rigidity constraint
always has been understand in a classical fashion without catching the
general structure underlying it.
In next sections I will show how Coriolis interaction and other features
of polyatomic  molecular spectra may be described
by means of geometry induced abelian and non-abelian monopole
gauge fields and quantum potentials appearing in the effective rotational
dynamics.

 For the moment let me prepare the way by adapting coordinates to $SO(3)$.
We introduce a cartesian frame $\x\y\z$ fixed in the
equilibrium configuration of the molecule. For sake of simplicity we shall
choose it as a principal frame  of inertia.  The unit vectors ${\bf e}_{\x},
{\bf e}_{\y},{\bf e}_{\z}$ along ${\x},{\y}$ and ${\z}$ will be expressed in
terms of the three Euler angles $\alpha,\beta,\gamma$  parametrizing the
rotational group $SO(3)$ and  the position of the $a-$th nucleus in the
${\x}{\y}{\z}$ frame is
specified by $\vec{\r}_a=(\x_a,\y_a,
\z_a)$, $a=1,...,N$,
which is thought as  function of the normal coordinates $\xi_i$,
$i=1,...,3N-6$, of the potential.
It is easy to be convinced that
$\{\alpha,\beta,\gamma,\xi_1,\xi_2,..., \xi_{3N-6}\}$
constitutes an {\sl adapted coordinates frame} in the sense specified
before. The transformation
$(\vec{r}_1,\vec{r}_2,...,\vec{r}_{N})\rightarrow
(\alpha,\beta,\gamma,\xi_1,\xi_2,...,\xi_{3N-6})$ takes the form
\begin{equation}
\vec{r}_a={\x}_a (\vec{\xi} ) {\bf e}_{{\x}}
              +{\y}_a (\vec{\xi} ) {\bf e}_{{\y}}
              +{\z}_a (\vec{\xi} ) {\bf e}_{{\z}}
\end{equation}
for $a=1,...,N$. Treating the constraint as in section $3$ we obtain the
effective rotational dynamics to be described by Hamiltonian (\ref{Heff}),
which is completely characterized by the geometric quantities describing the
embedding of $SO(3)$ in $R^{3N-3}$.
The indexes $\mu,\nu$ run now over $1$ to $3$ and label the three Euler
angles. $g^{\mu\nu}$ is the inverse of the rigid body metric on
$SO(3)$
\be
&&g_{11}=(I_{{\x}}\cos^2\gamma +I_{{\y}}\sin^2\gamma)\sin^2\beta+
        I_{{\z}}\cos^2\beta,\nonumber\\
&&g_{12}=g_{21}=(I_{{\y}}-I_{{\x}})\sin\beta\sin\gamma\cos\gamma,
            \nonumber\\
&&g_{13}=g_{31}=I_{{\z}}\cos\beta,\nonumber\\
&&g_{22}=I_{{\x}}\sin^2\gamma+I_{{\y}}\cos^2\gamma,\nonumber\\
&&g_{23}=g_{32}=0\nonumber,\\
&&g_{33}=I_{{\z}},\nonumber
\ee
where $I_{{\x}},I_{{\y}},I_{{\z}}$ are the principal moments of
inertia of the molecule. The explicit form of the normal fundamental form
$A_\mu^{ij}$ depends on the system under consideration and
will be given in the sequel for some classes of molecules.
Finally for every electronic configuration the potential
$Q(\alpha,\beta,\gamma)$ results to be a constant expressed in terms of the
inertia tensor $I$ as
\be
Q=\hbar^2{\mbox{tr}\ (I^2)- (\mbox{tr}\ I)^2\over 2\ \mbox{det}\ I}.
\ee

A quick look to Eqs.(\ref{Heff}), (\ref{gf}) and (\ref{Qb}) illustrates very
clearly and concisely how rotational motion couples with molecular
vibrations. Effects connected with geometry induced gauge fields  are
expected for molecules presenting degenerate proper frequencies,
while the quantum potential (\ref{Qb}) produces a rotovibrational coupling even
in the non-symmetrical case. To appreciate the difference with the standard
treatment of the rigidity constraint compare Hamiltonian (\ref{Heff}) with
Eq.$(III.2)$ of the classical report Ref.\cite{Nie}.
The gauge field (\ref{gf}) and the quantum potential (\ref{Qb}) are neglected
in the rotational Hamiltonian and considered only as  perturbative
corrections.

\

\section[Monopoles, Quantum Potentials and the $XY_2$ molecule]{Monopoles,
         Quantum Potentials \\ and the linear symmetric $XY_2$ molecule}
\

 Let me start by briefly recalling the semiclassical picture describing the
rotovibrational spectrum of the linear symmetric $XY_2$ molecule. In first
approximation $XY_2$ is thought as a rigid dumbbell capable to perform small
vibrations around its equilibrium position and to rotate around
its body center. Rotational energies are described by that of a
spherical top
\be
E^{rot}={\hbar^2\over 2I}l(l+1),
\ee
$I$ being the momentum of inertia. The quantum number $l$ may in
principle assume every positive integer values $l=0,1,...$ . Molecular
vibrations may be described in terms of four uncoupled harmonic oscillator
by adapting normal coordinates.
Since $\omega_2$ is doubly degenerate
vibrational energies  may be written in terms of three quantum numbers
$n_1,n_2,n_3=0,1,...$ as
\be
E^{vib}=\hbar\omega_1(n_1+1/2)+\hbar\omega_2(n_2+1)+\hbar\omega_3(n_3+1/2).
\ee
The $XY_2$ spectrum is then expected to be described in rough approximation
by the sum of vibrational and rotational energies
\be
E=E^{vib}+E^{rot}
\ee
and in fact, apart from the absence of some  values of $l$, it is so.
In order to justify the absences is necessary to remember that in a many-body
system is not possible in general  to completely separate rotational and
vibrational degrees of freedom. When normal modes corresponding to the
degenerate frequency $\omega_2$ are both exited the molecule acquires a purely
vibrational angular momentum $l^{vib}$ so that in the
vibrational
state in which $l^{vib}$ is acting, the total angular momentum should be
quantized as
\be
l=|l^{vib}|,|l^{vib}|+1,... \nn
\ee
in accordance with observed spectra. For a more accurate description is then
necessary to consider anharmonicity and further rotovibrational interactions.

Now I come to the general analysis of the $XY_2$ molecule. Chosen the
$\x\y\z$ frame with the $\z$ axis coinciding with
the molecular axis, the orientation of the molecule does not  depend on the
$\gamma$ angle and the rotational configuration space reduces from $SO(3)$ to
the sphere $S^2$. The Euler angles $\alpha,\beta$ coincide then with the
standard $\phi,\theta$ spherical angles. The investigation of the $XY_2$
molecule is therefore reduced to the study of the immersion of the sphere $S^2$
in the six-dimesional euclidean space $R^6$. The rigid body metric on $S^2$ is
given by
\be
g_{\mu\nu} = \pmatrix{
            I\sin^2\beta & 0 \cr
            0 &I}, \label{ms}
\ee
Denoted by $\xi_1,\xi_{2a},\xi_{2b},\xi_3$ the normal vibrational coordinates,
the only nonvanishing components of second fundamental form an normal
fundamental form are
\be
\alpha^{1}_{\mu\nu}=\pmatrix{
            -\sqrt{I}\sin^2\beta & 0 \cr
            0 & - \sqrt{I}}
\ee
and
\be
 &&A_\mu^{2a2b}=(\cos\beta,0),\nn\\
 &&A_\mu^{2a3} =(\sin\beta,0),\label{nffs}\\
 &&A_\mu^{2b3}=(0,1).\nn
\ee
These quantities characterize the rotovibrational dynamics of the molecule
completely \cite{Mar}. For what concern the gauge field (\ref{gf})
 only the term $A_{\mu}^{2a2b}$
is relevant. The effective rotational dynamics is described by  Hamiltonian
(\ref{Heff}) with metric (\ref{ms}) and $A^{ij}_\mu$ given by (\ref{nffs}).
The motion on $S^2$ results then equivalent to that of a  charged particle
in a background magnetic-monopole field \cite{Dir,WY}. Chosen the eigenvalues
basis of $H^{(0)}$ is such a way that also $L_{2a2b}$ is diagonal,
the monopole charge is proportional to the angular momentum of the degenerate
oscillator $\langle L_{2a2b}\rangle$, that is to the vibrational angular
momentum $l^{vib}$. $l^{vib}$  is  always an integer. To discuss the motion in
the monopole background it is convenient to introduce the angular momentum
operators
\be
\begin{array}{l}
\ds\tilde{L}_1 =i(\cos\alpha \cot\beta\partial_\alpha+\sin\alpha\partial_\beta)
             +l^{vib}{\cos\alpha\over\sin\beta},\\
\ds\tilde{L}_2 =i(\sin\alpha \cot\beta\partial_\alpha-\cos\alpha\partial_\beta)
             +l^{vib}{\sin\alpha\over\sin\beta},\\
\ds\tilde{L}_3 =-i\partial_\alpha,
\end{array}
\ee
which close in the $SU(2)$ algebra. Then the rotational Hamiltonian
(\ref{Heff}) coincide , up to the ${\bar Q}$ potential, with that of a
spherical top
\be
\H_{rot}^{(n_1,n_2,n_3)}={\hbar^2\over 2I} (\tilde{L}^2-{l^{vib}}^2),
\label{Hl}
\ee
the only effect of the monopole being that of making the rotational quantum
number $l$ to take the values $l=|l^{vib}|,|l^{vib}|+1,...$ \cite{Dir,WY}.
The monopole gauge field induced by the reduction of the motion from $R^6$
to $S^2$ takes therefore into account the absence of the first $|l^{vib}|$
values of the rotational quantum number $l$. This phenomenon is very close to
that discovered by J.\ Moody, A.\ Shpere and F.\ Wilczek in the geometric phase
analysis of diatoms \cite{MSW}.

 To complete the analysis of the $XY_2$ spectrum at this order, cubic and
quartic terms of the Born-Oppenheimer potential as well as the quantum
potential ${\bar Q}$ have also to be taken into account. As well known
anharmonicity contributes to the spectrum the correction
\be
\Delta = 2\pi\hbar c \left(\sum_{p<q=1}^{3}x_{pq}
                     \left(n_p+d_p/2\right)
                     \left(n_q+d_q/2\right)
                      +y l^2\right)
\ee
(recall $d_1=d_3=1,d_2=2$).
Let us therefore concentrate on the potential ${\bar Q}$. Considered the
explicit expression of the metric (\ref{ms}) and of the normal fundamental form
(\ref{nffs}) a straightforward computation produces
\be
{\bar Q}
={1\over 2I}\left(\langle L_{2a3}^2\rangle+\langle L_{2b3}^2\rangle\right),
\label{Qxy2}
\ee
where the angled brakets denote expectation values in
states with the same energy. The matrix potential (\ref{Qxy2}) does not  depend
on rotational coordinates and results to be diagonal in every $H^{(0)}$'s
eigenvalues basis. It contributes to the spectrum the correction
\be
\Delta^{{\bar Q}}={\hbar^2\over 2I}
               \left({\omega_2\over\omega_3}+{\omega_3\over\omega_2}\right)
               \left(n_2+1\right)
               \left(n_3+1/2\right)
\label{Deltaxy2}
\ee
As an example for the $CO_2$ molecule $x_{23}\simeq -11$ {\it cm}$^{-1}$
whereas the coefficient $\hbar(\omega_2/\omega_3+\omega_3/\omega_2)/4I \pi c$
may
be estimated as $\simeq 2$ {\it cm}$^{-1}$. Corrections of this kind have been
foreseen in Ref.\cite{Her}. The analysis of the rigidity constraint proposed
in this paper shades light on their geometrical nature giving a very general
and compact formula to compute them.

\vskip2cm

\section{Monopoles and Symmetric Tops}
\

The Coriolis interaction between vibrational and rotational degrees of freedom
in a linear molecule are taken into account by a background monopole field and
a quantum potential acting on the rotational configuration space, the sphere
$S^2$. In this and in the next section I consider the case in which the whole
group $SO(3)$ is involved in the rotational dynamics, investigating the
geometrical nature of Coriolis interactions in a symmetric top and a spherical
top molecule. The problem requires the analysis of the embedding of $SO(3)$ in
an appropriate $n$-dimesional euclidean space.

 The simpler example of a symmetric top having degenerate frequencies is the
equilateral $X_3$ molecule. Chosen the frame $\x\y\z$ with the molecule in the
$\x\y$ plane and denoted by $2I$ the momentum of inertia relative to the $\z$
axis, the rigid body metric on $SO(3)$ reads
\be
g_{\mu\nu} = \pmatrix{
            I(1+\cos^2\beta) & 0 & 2I\cos\beta\cr
            0 & I & 0\cr
            2I\cos\beta & 0 & 2I}. \label{msyt}
\ee
$X_3$ posses one simple, $\omega_1$, and one double, $\omega_2$, proper
frequencies and I will denote by $\xi_1,\xi_{2a},\xi_{2b}$ the
corresponding vibrational coordinates. The second fundamental form and normal
fundamental form of the embedding of $SO(3)$ in $R^6$ may easily be evaluated
and result different from zero. For what concern the effective rotational
motion only the normal fundamental form is relevant. Its nonvanishing
component is given by
\be
A_\mu^{2a2b}=(\cos\beta,0,0).
\label{gfsyt}
\ee
The rotational dynamics is described by Hamiltonian (\ref{Heff}) with
$g_{\mu\nu}$ given by (\ref{msyt}) and $A^{ij}_\mu$ by (\ref{gfsyt}).
Chosen an eigenvalues basis of $H^{(0)}$ such that $L_{2a2b}$ is diagonal and
denoted again with $l^{vib}=\langle L_{2a2b}\rangle$ the vibrational angular
momentum, the motion on $SO(3)$ take place in presence of a background
monopole-like field with charge $l^{vib}$. To discuss the $SO(3)$ dynamics
in presence of the monopole is again possible to introduce adapted angular
momentum  operators
\be
\begin{array}{l}
\ds\tilde{L}_1 =i\left(
                {\cos\gamma\over \sin\beta} \partial_\alpha
                - \sin\gamma                 \partial_\beta
                - \cot\beta\cos\gamma        \partial_\gamma
               \right)
                -l^{vib}\cot\beta\cos\gamma,\\
\ds\tilde{L}_2 =i\left(
                 {\sin\gamma\over \sin\beta} \partial_\alpha
                + \cos\gamma                 \partial_\beta
                - \cot\beta\sin\gamma        \partial_\gamma
               \right)
                -l^{vib}\cot\beta\sin\gamma,\\
\ds\tilde{L}_3 = -i\partial_\gamma-l^{vib},
\end{array}
\ee
which close in the $SU(2)$ algebra. Hamiltonian (\ref{Heff}) for the $X_3$
molecule takes then the standard form
\be
H_{rot}^{(n_1,n_2)}= {\hbar^2\over 2I}(\tilde{L}_1^2+\tilde{L}_2^2)
                    +{\hbar^2\over 4I}(\tilde{L}_3+l^{vib})^2.
\label{Hsyt}
\ee
introduced in Ref.\cite{TJD} to take account of Coriolis coupling.
The   monopole field induced by the rigidity constraint
in the effective rotational hamiltonian  is therefore responsible
for Coriolis interactions between vibrational and rotational degrees of
freedom. The quantum potential (\ref{Qb}) does not  play any role in the
description of the spectrum being identically zero.

\

\section[Non-Abelian Monopoles and Spherical Tops]{Non-Abelian Monopoles
         \\ and Spherical Tops}
\

If a molecule has the symmetry of one of the cubical groups its proper
frequencies may be triply degenerate and the gauge field appearing in the
effective rotational Hamiltonian non-abelian. This is the case of spherical top
molecules. The simpler example of physical interest is the tetrahedral $XY_4$
molecule. It possesses one simple, $\omega_1$, one double, $\omega_2$,
and two triple, $\omega_3$, $\omega_4$, proper frequencies
\cite{Her}. For sake of clearness I will however consider the simpler example
of a tetrahedral $Y_4$ molecule in which the effect I want to point out
already appears. $Y_4$ may be thought as an $XY_4$ in which the second triple
degenerate normal mode may be neglected. Chosen the $\x\y\z$ frame fixed in
the equilibrium configuration of the molecule in such a way that the four atoms
lie in the positions
$({\lambda\over\sqrt{8}},{\lambda\over\sqrt{8}},{\lambda\over\sqrt{8}})$,
$(-{\lambda\over\sqrt{8}},-{\lambda\over\sqrt{8}},{\lambda\over\sqrt{8}})$,
$({\lambda\over\sqrt{8}},-{\lambda\over\sqrt{8}},-{\lambda\over\sqrt{8}})$,
$(-{\lambda\over\sqrt{8}},{\lambda\over\sqrt{8}},-{\lambda\over\sqrt{8}})$
and denoted by $I$ the momentum of inertia relative to the axis $\x$,$\y$ and
$\z$, the rigid body metric on $SO(3)$ reads
\be
g_{\mu\nu} = \pmatrix{
            I & 0 & I\cos\beta\cr
            0 & I & 0\cr
            I\cos\beta & 0 & I}. \label{mspt}
\ee
The second fundamental form and normal fundamental form of the embedding of
$SO(3)$ in $R^9$ may also be easily evaluated. Denoting as before with
$\xi_1,\xi_{2a},\xi_{2b},\xi_{3a},\xi_{3b},\xi_{3c}$ the normal coordinates,
the
nonvanishing components of the normal fundamental form reads
\be
&&A_\mu^{2a3a}=\left(-{1+\sqrt{3}\over 2\sqrt{2}}\sin\beta\sin\gamma,
              -{1+\sqrt{3}\over 2\sqrt{2}}\cos\gamma,0\right),\nonumber\\
&&A_\mu^{2a3b}=\left( {1-\sqrt{3}\over 2\sqrt{2}}\sin\beta\cos\gamma,
              -{1-\sqrt{3}\over 2\sqrt{2}}\sin\gamma,0\right),\nonumber\\
&&A_\mu^{2a3c}=\left( {1\over\sqrt{2}}\cos\beta,
               0,{1\over\sqrt{2}}\right),\nonumber\\
&&A_\mu^{2b3a}=\left( {1-\sqrt{3}\over 2\sqrt{2}}\sin\beta\sin\gamma,
               {1-\sqrt{3}\over 2\sqrt{2}}\cos\gamma,0\right),\nonumber\\
&&A_\mu^{2b3b}=\left(-{1+\sqrt{3}\over 2\sqrt{2}}\sin\beta\cos\gamma,
               {1+\sqrt{3}\over 2\sqrt{2}}\sin\gamma,0\right),\label{gfspt}\\
&&A_\mu^{2b3c}=\left(-{1\over\sqrt{2}}\cos\beta,
               0,-{1\over\sqrt{2}}\right)\nonumber\\
\mbox{and}\ \ \ \  &&  \nonumber\\
&&A_\mu^{3a3b}=\left(-{1\over 2}\cos\beta,0,-{1\over 2}\right),\nonumber\\
&&A_\mu^{3b3c}=\left(-{1\over 2}\sin\beta\sin\gamma,
              -{1\over 2}\cos\gamma,0\right),\nn\\
&&A_\mu^{3c3a}=\left( {1\over 2}\sin\beta\cos\gamma,
              -{1\over 2}\sin\gamma,0\right).\nonumber
\ee
The effective rotational motion is again described by Hamiltonian (\ref{Heff})
with $g_{\mu\nu}$ the spherical top metric (\ref{mspt}) and  $A^{ij}_\mu$
given by (\ref{gfspt}).
The vibrational angular momentum is not in general diagonalizable.
Nevertheless in every vibrational state labelled by the
quantum numbers $n_1,n_2$ and $n_3$, the matrices
$L^{vib}_1=\langle L_{3c3a}\rangle/\hbar$,
$L^{vib}_2=\langle L_{3c3b}\rangle/\hbar$ and
$L^{vib}_3=\langle L_{3a3b}\rangle/\hbar$ form a
$n_2(n_2+1)(n_3+1)(n_3+2)/4$-dimensional
representation of the Lie algebra of $SO(3)$ so that $A_\mu$ is an
$SO(3)$ gauge potential presenting the characteristic of a  non-abelian
monopole. Stated in other words we are facing the dynamical problem of the
motion on the group $SO(3)$ in presence of a non-abelian monopole
whose gauge
group is again $SO(3)$. Once more we address the solution by considering
appropriate angular momentum operators which close in the $SU(2)$ algebra
\be
\begin{array}{l}
\ds\tilde{L}_1 =i\left(
                {\cos\gamma\over \sin\beta} \partial_\alpha
                - \sin\gamma                 \partial_\beta
                - \cot\beta\cos\gamma        \partial_\gamma
               \right),\\
\ds\tilde{L}_2 =i\left(
                 {\sin\gamma\over \sin\beta} \partial_\alpha
                + \cos\gamma                 \partial_\beta
                - \cot\beta\sin\gamma        \partial_\gamma
               \right),\\
\ds\tilde{L}_3 = -i\partial_\gamma.
\end{array}
\ee
Introduced the vector matrix $L^{vib}=(L^{vib}_1,L^{vib}_2,L^{vib}_3)$ and
denoting by $\I$ the $n_2(n_2+1)(n_3+1)(n_3+2)/4$-dimensional identity matrix,
Hamiltonian (\ref{Heff}) for the $Y_4$ molecule assume the standard form
\cite{TJD}
\be
\H_{rot}^{(n_1,n_2,n_3)}= {\hbar^2\over 2I}
          \left({\I}\tilde{L}-{1\over 2}L^{vib}\right)^2,
\label{Hspt}
\ee
up to the ${\bar Q}$ potential.
 In a spherical top molecule Coriolis interactions are therefore described by
means of a non-abelian monopole gauge field induced in the effective rotational
dynamics  by the rigidity constraint.

The quantum potential ${\bar Q}$ plays also an important role in the
description of rotovibrational interactions.  The use of the explicit form of
the metric (\ref{mspt}) and of the normal fundamental form (\ref{gfspt})
allows to rewrite the general formula (\ref{Qb})  as
\be
{\bar Q}&=&{\hbar^2\over 8I}
 \left({\over}\langle L_{2a3a}^2+L_{2a3b}^2+L_{2a3c}^2
              +L_{2b3a}^2+L_{2b3b}^2+L_{2b3c}^2\rangle+\right.\nn\\
 &&   \ \ \ \ \    +\sqrt{3}\langle L_{2a3a}^2-L_{2b3a}^2 \rangle
         -\sqrt{3}\langle L_{2a3b}^2-L_{2b3b}^2 \rangle + \nn\\
 &&   \ \ \ \ \    + \langle L_{2a3a}L_{2b3a}+L_{2b3a}L_{2a3a} \rangle \nn\\
 &&   \ \ \ \ \    + \langle L_{2a3b}L_{2b3b}+L_{2b3b}L_{2a3b} \rangle \nn\\
 &&   \ \ \ \ \
\left. -2\langle L_{2a3c}L_{2b3c}+L_{2b3c}L_{2a3c}\rangle {\over}\right).\nn
\ee
To evaluate the expectation values is convenient to rewrite vibrational angular
momentum operators in terms of destruction and creation operators
\be
\left\{ \begin{array}{l}
a_i=\sqrt{\omega_i\over2\hbar}
      \left(\xi_i+{1\over\omega_i}{\partial\over\partial\xi_i}\right)\\
a^{\dagger}_i=\sqrt{\omega_i\over2\hbar}
      \left(\xi_i-{1\over\omega_i}{\partial\over\partial\xi_i}\right)
       \end{array}\right.
\ee
for $i=2a,2b,3a,3b,3c$. Computation conduces then to the expression
\be
{\bar Q}&=&{\hbar^2\over 8I}
   \left({\omega_3\over\omega_2}+{\omega_2\over\omega_3}\right)\times \nn\\
&&\left\langle 2(a^{\dagger}_{2a}a_{2a}+a^{\dagger}_{2b}a_{2b}+1)
(a^{\dagger}_{3a}a_{3a}+a^{\dagger}_{3b}a_{3b}+a^{\dagger}_{3c}a_{3c}+3/2)+
 \right. \nn\\
&&+(a_{2a}a^{\dagger}_{2b}+a^{\dagger}_{2a}a_{2b})
     (a^{\dagger}_{3a}a_{3a}+a^{\dagger}_{3b}a_{3b}-2a^{\dagger}_{3c}a_{3c})+
    \label{Qbx4}\\
&&\left. \sqrt{3}(a^{\dagger}_{2a}a_{2a}-a^{\dagger}_{2b}a_{2b})
                   (a^{\dagger}_{3a}a_{3a}-a^{\dagger}_{3b}a_{3b})\right
         \rangle, \nn
\ee
which is not diagonal in the standard basis of $H^{(0)}$. Nevertheless
$\left[H^{(0)},{\bar Q}\right]=0$ so that the diagonalization of ${\bar Q}$
is possible. We achieve this aim by performing a rigid gauge transformation
in the $\xi_{2a},\xi_{2b}$ subspace in correspondence of every
$n_{3a},n_{3b},n_{3c}$, ($n_i$ denotes, of course, the quantum number created
by $a^{\dagger}_i$; $n_2=n_{2a}+n_{2b}$, $n_3=n_{3a}+n_{3b}+n_{3c}$), that is
by
rotating creation-destruction operators by
\be
\left\{\begin{array}{l}
 a_{+} = \cos\psi\  a_{2a} +\sin\psi\  a_{2b}  \\
 a_{-} =-\sin\psi\  a_{2a} +\cos\psi\  a_{2b}
\end{array} \right.
\ee
where
\be
\cos\psi=\sqrt{{1\over2}+{\sqrt{3}\over4}
{n_{3a}-n_{3b}\over\sqrt{n_{3a}^2+n_{3b}^2+n_{3c}^2
                        -n_{3a}n_{3b}-n_{3b}n_{3c}-n_{3c}n_{3a}}}}\nn\\
\sin\psi=\sqrt{{1\over2}-{\sqrt{3}\over4}
{n_{3a}-n_{3b}\over\sqrt{n_{3a}^2+n_{3b}^2+n_{3c}^2
                        -n_{3a}n_{3b}-n_{3b}n_{3c}-n_{3c}n_{3a}}}}\nn
\ee
Rewriting (\ref{Qbx4}) in terms of $a_{+},a^{\dagger}_{+},a_{-},
a^{\dagger}_{-}$ and
denoting by $n_{+}, n_{-}$ ($n_2=n_{+}+n_{-}$) the corresponding quantum
numbers, we obtain the correction to the spectrum
\be
\begin{array}{rl}
\Delta^{{\bar Q}}&= \ds{\hbar^2\over 4I}
  \left({\omega_3\over\omega_2}+{\omega_2\over\omega_3}\right)
\left[{\over}(n_2+1)(n_3+3/2)\right.+\\
&\ds+\left.2(n_{+}-n_{-})\sqrt{n_{3a}^2+n_{3b}^2+n_{3c}^2
                        -n_{3a}n_{3b}-n_{3b}n_{3c}-n_{3c}n_{3a}}\right].
\end{array}
\label{Deltax4}
\ee
 ${\bar Q}$ contributes therefore in removing the vibrational degeneracy
together with anharmonic terms of the Born-Oppenheimer potential.

 As for the $XY_2$ and the $X_3$ molecules, gauge fields and quantum potential
induced by the reduction of the quantum dynamics from $R^9$ to $SO(3)$ take
naturally into account rotovibrational interactions giving a general and simple
way to compute them.

\

\section{Concluding Remarks}
\

 In this paper I discussed the constrained motion of a quantomechanical system
by explicitly considering the mechanism realizing the constraint. This gives
rise to a very rich dynamical structure unrecoverable with the standard
techniques.
I focused my attention on the first order term of the perturbative expansion
(\ref{exp}), showing how gauge fields and quantum potentials induced in the
effective dynamics are responsible for leading order rotovibrational
interactions in polyatomic molecules. The main results of the paper are
\begin{description}
\item{---} The interpretation of the standard Hamiltonians (\ref{Hl}),
(\ref{Hsyt}) and (\ref{Hspt}), describing Coriolis interactions in linear,
symmetric top and spherical top molecules, in terms of the general Hamiltonian
(\ref{Heff}). This illustrates that the gauge field (\ref{gf}) is relevant in
the description of constrained systems, giving an explicit realization in
nature
of abelian and non-abelian monopole fields.
\item{---} The very general and compact expression (\ref{Qb}) to compute
corrections
to molecular spectra given by the other  leading order rotovibrational
interactions, such as (\ref{Deltaxy2}) for the $XY_2$ molecule and
(\ref{Deltax4}) for the $Y_4$ molecule.
\item{---} A complete geometrical characterization of molecular rotovibrational
dynamics, showing how the intrinsic and extrinsic geometrical properties of the
embedding of the rotational group $SO(3)$ in the configuration space $R^n$,
 together with the Born-Oppenheimer potential, completely determine the
rotovibrational spectrum of a polyatomic molecule.
\end{description}

 It is possible to demonstrate that the whole perturbative expansion
(\ref{exp})
is completely characterized by $g_{\mu\nu}$, $\alpha^{i}_{\mu\nu}$ and
$A^{ij}_{\mu}$ giving the explicit expression of the arbitrary order term
\cite{Mar}. The method is therefore suitable for a complete perturbative
analysis of the spectra of rigid molecules. I want to point out that this
does not  constitute an alternative to the standard techniques \cite{Nie},
but, rather, an improvement in the light of the general considerations
of section 3 and Refs.\cite{MD,FO,Mar}. Besides the problem of taking into
account finer corrections of rigid molecular spectra by considering further
terms of expansion (\ref{exp}), several other questions remain open.
It is possible, for example, to correlate the form of the monopole gauge field
appearing in the effective rotational dynamics to the symmetry of the molecule,
is such a way that effective Hamiltonians for more complex molecules may be
written automatically?
It is possible, by resumming part of expansion
(\ref{exp}), to take into account non-rigidity effects in molecular spectra?
More in general it would be interesting to investigate the relation between
constrained quantum mechanics and geometric phases.

 To conclude I want to point out the very important fact that the
perturbative expansion (\ref{exp}) for molecular systems writes naturally in
terms of modified angular momentum operators and creation/destruction
operators corresponding to normal modes, so that the evaluation of molecular
spectra and eigenfunctions may be performed completely by means of algebraic
manipulations \cite{Iac}.

\

\section*{Aknoledgments}

I wish to warmly thank C.\ Destri and E.\ Onofri for useful discussions.

\end{document}